\documentclass{IEEEtran}
\usepackage{amssymb}
\usepackage{amsfonts}
\usepackage{amsmath}
\usepackage{algpseudocode}
\usepackage{algorithm}
\usepackage{caption}
\usepackage{bm}
\usepackage{graphicx,subfigure,cite,multicol,multirow,diagbox,booktabs,array}
\usepackage{stfloats}

\usepackage{color}
\usepackage{float}
\usepackage{subfig}
\ifCLASSINFOpdf
\else
\fi
\begin{document}
\title{Machine Learning Methods for Inferring the Number of UAV Emitters via Massive MIMO Receive Array}
\author{Yifan Li, Feng Shu,~\emph{Member},~\emph{IEEE}, Jinsong Hu, Shihao Yan, Haiwei Song, Weiqiang Zhu, Da Tian, Yaoliang Song,~\emph{Senior Member},~\emph{IEEE}, and Jiangzhou Wang,~\emph{Fellow},~\emph{IEEE}
\thanks{Y. Li, and Y. Song are with the School of Electronic and Optical Engineering, Nanjing University of Science and Technology, Nanjing 210094, China (e-mail: liyifan97@foxmail.com).}
\thanks{F. Shu is with the School of Information and Communication Engineering, Hainan University, Haikou 570228, China (e-mail: shufeng0101@163.com).}
\thanks{Jinsong Hu is with the College of Physics and Information Engineering, Fuzhou University, Fuzhou, Fujian, 350116, China (e-mail: jinsong.hu@fzu.edu.cn).}
\thanks{S. Yan is with the School of Science and Security Research Institute, Edith Cowan University, Perth, WA 6027, Australia (e-mail: s.yan@ecu.edu.au).}
\thanks{H. Song, W. Zhu and D. Tian are with the 8511 Research Institute, China Aerospace Science and Industry Corporation, Nanjing 210007, China (e-mail: hw8511@126.com).}
\thanks{J. Wang is with the School of Engineering, University of Kent, Canterbury CT2 7NT, U.K (e-mail: j.z.wang@kent.ac.uk).}
}\maketitle

\begin{abstract}
To provide important prior knowledge for the DOA estimation of UAV emitters in future wireless networks, we present a complete DOA preprocessing system for inferring the number of emitters via massive MIMO receive array. Firstly, in order to eliminate the noise signals, two high-precision signal detectors, square root of maximum eigenvalue times minimum eigenvalue (SR-MME) and geometric mean (GM), are proposed. Compared to other detectors, SR-MME and GM can achieve a high detection probability while maintaining extremely low false alarm probability. Secondly, if the existence of emitters is determined by detectors, we need to further confirm their number. Therefore, we perform feature extraction on the the eigenvalue sequence of sample covariance matrix to construct feature vector and innovatively propose a multi-layer neural network (ML-NN). Additionally, the support vector machine (SVM), and naive Bayesian classifier (NBC) are also designed. The simulation results show that the machine learning-based methods can achieve good results in signal classification, especially neural networks, which can always maintain the classification accuracy above 70\% with massive MIMO receive array. Finally, we analyze the classical signal classification methods, Akaike (AIC) and Minimum description length (MDL). It is concluded that the two methods are not suitable for scenarios with massive MIMO arrays, and they also have much worse performance than machine learning-based classifiers.
\end{abstract}
\section{Introduction}
With the advantages of high mobility and low cost, unmanned aerial vehicles (UAVs) are always supposed to play important roles in wireless networks for implementing the tasks like weather monitoring, traffic control, emergency search, communication relaying, etc. \cite{zeng2016wireless}. However, different from the traditional ground-to-ground (G2G) communications, UAV communications have some special characteristics and challenges, e.g., the high mobility will lead to the UAV communication channels change much faster, the high flight altitude requiring the ground base stations to provide larger 3D signal coverage for UAVs, the line of sight (LoS) paths between UAVs and base stations are vulnerable to interference from ground users over the same frequency \cite{huang2021massive}.
Obviously, 4G wireless networks are difficult to meet the requirements for UAV communications. But as is known to us, massive multiple-input multiple-output (MIMO) is a key technology in 5G or future 6G systems \cite{wang2014cellular}, \cite{saad2019vision}, it can make a significant improvements in system capacity, reliability, and spectral efficiency by using techniques like spatial multiplexing, diversity, and beamforming \cite{zhang20196g}. Compared to small arrays, the higher array gain of massive MIMO arrays can make a great extension of signal coverage \cite{chandhar2019massive}, and experiment results in \cite{harris2017performance} showed massive MIMO works well with LoS mobile channels.
So in view of the problems that UAV communications faced, it is natural to consider the combination of UAVs and massive MIMIO technology \cite{geraci2022will}. In \cite{bai2022non}, a nonstationary 3D geometry-based model was proposed for UAV-to-ground massive MIMO channels, this model considered the realistic scenarios and discussed the impact of some important UAV parameters like altitude, flight velocity, so it can give some inspiration for the future research of the 6G standard UAV channel models. As UAVs often appear as clusters, the potential of massive MIMO ground station communicates with UAV swarms was explored in \cite{chandhar2017massive}, and a realistic geometric model was also developed.

Since the high mobility of UAVs, it is necessary for ground base stations to obtain direction-of-arrival (DOA) information of UAVs timely for channel estimation and communication security. For most DOA estimation algorithms, like MUSIC and ESPRIT, the number of emitters is a required prior knowledge, but the number is usually unknown \cite{huang2015source}. So the inferring the number of emitters has been an active topic in array processing for a few decades \cite{krim1996two}.
In recent years, the potential of massive MIMO technology in array processing has also been gradually discovered, for the larger number of antennas can decrease the beamwidth and then increase the angular resolution of the arrays \cite{bjornson2019massive}.    
Therefore, considering the realistic needs of UAV communications, and the advantages of massive MIMO technology in array processing, we will study the methods for inferring the number of UAV emitters via massive MIMO receive array in this work.

In general, the solutions for inferring the number of emitters can be divided into two main categories, the first is based on the information theoretic criteria and another is based on the analysis of the covariance matrices. Since detecting the number of signal sources can be viewed as a typical model order selection problem, Akaike firstly proposed a method focusing on finding the minimum Kullback-Leibler (KL) discrepancy between the probability density function (PDF) of obtained data and that of models for selection\cite{akaike1974new}, and this method is called AIC now. Schwarz introduced Bayesian information criterion (BIC) based on Akaike's work\cite{schwarz1978estimating}, and Rissanen also derived a similar criterion called MDL\cite{rissanen1978modeling}. \cite{stoica2004model} provided a good summary of these classical information criteria. In the last decade, Lu and Zoubir proposed the generalized Bayesian information criterion (GBIC)\cite{lu2012generalized} and flexible detection criterion (FDC)\cite{lu2012flexible}, which effectively improved the performance on source enumeration.
The other basic method for enumerating the number of sources is performing analysis on the covariance matrices of signals received by arrays. Williams and Johnson proposed sphericity test for source enumeration in \cite{williams1990using}, which was based on a hypothesis test for the covariance matrix. \cite{brcich2002detection} gave a bootstrap-based method to estimate the null distributions of the test statistics. Wax and Adler solved this problem by performing signal subspace matching\cite{wax2021detection}.

Signal detection is another technique adopted in this work. In order to reduce the interference of the noise to the detection of signal number, some good methods were proposed such as classic signal detection algorithms containing energy detection\cite{cabric2004implementation}, matched-filter detection\cite{cabric2006spectrum}, cyclostationarity-based detection\cite{gardner1991exploitation}, etc.. On the basis of these methods, Zeng and Liang proposed two eigenvalue-based algorithms in \cite{zeng2009eigenvalue}, Zhang \emph{et al}. used the generalized likelihood ratio test (GLRT) approach to improve detection performance \cite{zhang2010multi} and an eigenvalue-based LRT algorithm was also given in \cite{liu2016optimal}.

Machine learning (ML) has played an important role in the fields of signal processing and communications for many years \cite{anderson2004model}, and now the ML-based methods used in 5G mainly including supervised learning, unsupervised learning and reinforcement learning\cite{jiang2016machine}. Thilina \emph{et al}. compared the performance of unsupervised learning approaches and supervised learning approaches for cooperative spectrum sensing\cite{thilina2013machine}. A machine learning-based DOA measurement method was also proposed in \cite{zhuang2020machine}. And \cite{shu2021spatial} used neural network for power allocation in wireless communication network.

In this paper, we will combine the techniques mentioned above for inferring the number of UAV emitters via massive MIMO receive array. First, the pure noise signals are separated by threshold detectors, and then the feature vectors are extracted from the sample covariance matrices of the remaining signals. Finally, the ML-NN and other machine learning methods are used to classify the signals for determining the number of emitters. Therefore, our main contributions are summarized as follows:
\begin{enumerate}
	\item A DOA preprocessing system is proposed for obtaining the number of UAV emitters via a massive MIMO array. The main steps of this system include signal detection and inferring the number of emitters. The sampled baseband signal is first inputted into signal detectors. If detection result shows the presence of emitters, this signal is further transmitted to signal classifiers to determine the number of emitters.
	\item Two high-precision signal detectors, square root of maximum eigenvalue times minimum eigenvalue (SR-MME) and geometric mean (GM), are proposed in Section \ref{signal detectors}. Their thresholds and probability of detection are also derived with the aid of random matrix theories. The simulation results show that, SR-MME and GM have significant improvement in detection performance compared with MME detector proposed in \cite{zeng2009eigenvalue} and M-MME detector proposed in \cite{jie2021high}, even SNR is very low and number of samples is small. The simulation results also show that SR-MME and GM can maintain a low false alarm probability while achieving a high detection probability.
	\item Since the existence of emitters is known, we innovatively introduce machine learning-based classifiers to infer their number, including multi-layer neural networks (ML-NN), support vector machine (SVM), and naive Bayesian classifier (NBC). Important features which make up feature vectors are also extracted from eigenvalue sequences of signals' sample covariance matrices. The results show that machine learning methods are very suitable for performing signal classification, especially neural networks, because they can achieve a classification accuracy of 70\%, even under extreme conditions. Finally, we validate the classification performance of AIC and MDL under different SNR and number of receive antennas. We show that they are unapplicable to scenarios with low SNR and massive MIMO receive arrays compared to machine learning-based methods.
\end{enumerate}

The rest of the paper is organized as follows. In Section \ref{system model}, we present specific system model and assumptions. Two high precision signal detectors are given in Section \ref{signal detectors}. Section \ref{signal classifier} shows how to perform feature extraction on received signals and classify them by machine learning methods. Then, the advantages of proposed detectors and classifiers are presented through simulation results in Section \ref{simulation}. Finally, Section \ref{conclusion} draws conclusions.

\emph{\rm{\textbf{Notation}}:} Matrices, vectors, and scalars are denoted by letters of bold upper case, bold lower case, and lower case, respectively. Signs $(\cdot)^T$, $(\cdot)^\ast$ and $(\cdot)^H$ represent transpose, conjugate and conjugate transpose. $\mathbf{I}_M$ denotes the $M\times M$ identity matrix. $\rm{diag}\{\cdot\}$ stands for diagonal matrix.

\begin{figure}[ht]
  \centering
  \includegraphics[width=0.43\textwidth]{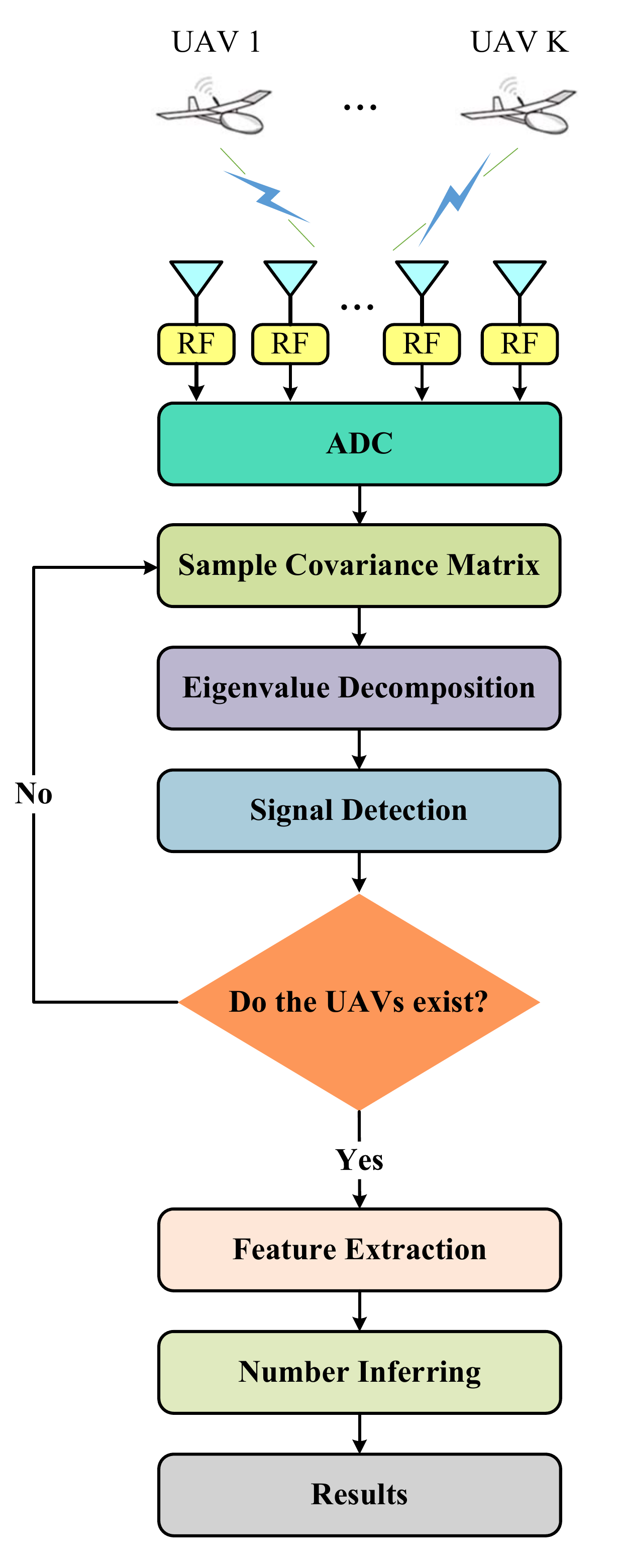}\\
  \caption{System flowchart.\label{system flowchart}}
\end{figure}

\section{System Model}\label{system model}
As the system shown in Fig.\ref{system flowchart}, we consider a scenario with $K$ far-field UAV emitters and one massive MIMO receiver equipped with an $M$-element uniform linear array (ULA). The signals transmitted by $k$th UAV is denoted by $s_k(t)e^{j2\pi f_c t}$, where $s_k(t)$ is baseband signal and $f_c$ is carrier frequency. Referring to \cite{zhang2021direction}, the received signals at the $m$th antenna is given by 
\begin{equation}
	y_m(t)=\sum_{k=1}^K s_k(t)e^{j2\pi f_c t}e^{-j2\pi f_c \tau_{k,m}}+v_m(t),
\end{equation}
where $v_m(t)\sim\mathcal{CN}(0,\sigma_{v}^2)$ represents the additive white Gaussian noise (AWGN) term, and $\tau_{k,m}$ denotes the propagation delay from the $k$th UAV to $m$th antenna is expressed by
\begin{equation}
	\tau_{k,m}=\tau_0-\frac{(m-1)d\sin \theta_k}{c},
\end{equation}
where $\tau_0$ is the propagation delay from the UAV to the reference point on the receive array, $\theta_k$ is the angle of signal incidence from the $k$th UAV, $d=\lambda/2$ represents the space between array elements and $c$ denotes speed of light. Then received signals go through ADC and down converter, we can get 
\begin{equation}
	y_m(n)=\sum_{k=1}^K e^{-j2\pi(m-1)d\sin\theta_k/\lambda}s_k(n)+v_m(n),
\end{equation}
and by combining all the $M$ antennas, we obtain
\begin{equation}
	\mathbf{y}(n)=\sum_{k=1}^K \mathbf{a}(\theta_k)s_k(n)+\mathbf{v}(n)\label{signal model},
\end{equation}
where $\mathbf{v}(n)=[v_1(n),\cdots,v_M(n)]^T$ denotes the noise vector and
\begin{equation}
	\mathbf{a}(\theta_k)=[1,e^{-j2\pi d\sin \theta_k/\lambda},\cdots,e^{-j2\pi (M-1)d\sin \theta_k/\lambda}]^T,
\end{equation}
is the array manifold.

Initially, it is not clear whether the UAVs exist, so we should consider two situations including the signals presence and only noise\cite{chen2008stochastic}. By turning (\ref{signal model}) to matrix form, we can get
\begin{equation}
	H_0:\mathbf{y}(n)=\mathbf{v}(n)~~
	H_1:\mathbf{y}(n)=\mathbf{A}\mathbf{s}(n)+\mathbf{v}(n),
\end{equation}
where $\mathbf{s}(n)=[s_1(n),\cdots,s_K(n)]^T$, $\mathbf{A}=[\mathbf{a}(\theta_1),\cdots,\mathbf{a}(\theta_K)]$. Then the covariance matrix of the received signal can be expressed by
\begin{equation}
	\mathbf{Q}_\mathbf{y}=\mathbf{A}\mathbf{Q}_{\mathbf{s}}\mathbf{A}^H+\sigma_{\mathbf{v}}^2\mathbf{I}_M=\sum_{k=1}^K\sigma_{\mathbf{s},k}^2\mathbf{a}(\theta_k)\mathbf{a}^H(\theta_k)+\sigma_{\mathbf{v}}^2\mathbf{I}_M.
\end{equation}
where $\mathbf{Q}_{\mathbf{s}}=\rm{E}[\mathbf{S}(n)\mathbf{S}^H(n)]$$={\rm{diag}}\{\sigma_{\mathbf{s},1}^2,\cdots,\sigma_{\mathbf{s},K}^2\}$.

Since the base station is equipped with a massive array, $M\gg K$ and ${\rm{rank}(A)}=K$. Then the eigenvalues of $\mathbf{Q}_\mathbf{y}$ satisfy the following properties
\begin{equation}
	\underbrace{\lambda_1\geq\lambda_2\geq\cdots\geq\lambda_K}_{\rm{signal~subspace}}>\underbrace{\lambda_{K+1}=\cdots=\lambda_M=\sigma_{\mathbf{v}}^2}_{\rm{noise~subspace}}~~,
	\label{eigenvalue_character}
\end{equation}
and
\begin{equation}
	\lambda_m=\rho_m+\sigma_{\mathbf{v}}^2~~,
\end{equation}
where $\rho_1\geq\cdots\geq\rho_K>\rho_{K+1}=\cdots=\rho_M=0$ are the eigenvalues of $\mathbf{A}\mathbf{Q}_{\mathbf{s}}\mathbf{A}^H$.

In practice, the covariance matrix of received signal $\mathbf{y}$ can't be obtained accurately. So the sample covariance matrix of received signal is usually used to approximate it
\begin{equation}
	\hat{\mathbf{Q}}_{\mathbf{y}}=\frac{1}{N}\sum_{n=1}^N\mathbf{y}(n)\mathbf{y}^{H}(n)=\frac{1}{N}\mathbf{Y}\mathbf{Y}^H,\label{sample_covariance_matrix}
\end{equation}
where
\begin{equation}
	H_0:\mathbf{Y}=\mathbf{V}~~~~
	H_1:\mathbf{Y}=
	\mathbf{A}\mathbf{S}+\mathbf{V}\label{two situations},
\end{equation}
and $\mathbf{S}=[\mathbf{s}(1),\mathbf{s}(2),\cdots,\mathbf{s}(N)]$, $\mathbf{V}=[\mathbf{v}(1),\mathbf{v}(2),\cdots,\mathbf{v}(N)]$.

\section{Signal Detectors}\label{signal detectors}
As shown in Fig.\ref{system flowchart}, after the sample covariance matrix of received signal is obtained, take eigenvalue decomposition (EVD) on it. For the two situations in (\ref{two situations}), eigenvalues are represented by $\lambda_1(\hat{\mathbf{Q}}_{\mathbf{y},H_0})\geq \cdots \geq \lambda_M(\hat{\mathbf{Q}}_{\mathbf{y},H_0})$ and $\lambda_1(\hat{\mathbf{Q}}_{\mathbf{y},H_1})\geq \cdots \geq \lambda_M(\hat{\mathbf{Q}}_{\mathbf{y},H_1})$ respectively. For convenience, we consider moving the constant $1/N$ to the left hand side of (\ref{sample_covariance_matrix}). Assuming $\sigma_{\mathbf{v}}^2=1$, we can get
\begin{subequations}
\begin{align}
&\mathbf{R}_{H_0}=\mathbf{V}\mathbf{V}^H,\\
&\mathbf{R}_{H_1}=N\mathbf{A}\hat{\mathbf{Q}}_\mathbf{S}\mathbf{A}^H+\mathbf{R}_{H_0},\label{R_H1}
\end{align}\label{matrix_R}
\end{subequations}
where $\mathbf{R}_{H_0}$ is a Wishart matrix and $\hat{\mathbf{Q}}_\mathbf{S}$ is sample covariance matrix of $\mathbf{S}$. The eigenvalues of $\mathbf{R}_{H_0}$ and $\mathbf{R}_{H_1}$ can also be expressed as $\lambda_1(\mathbf{R}_{H_0})\geq\cdots\geq\lambda_M(\mathbf{R}_{H_0})$ and $\lambda_1(\mathbf{R}_{H_1})\geq\cdots\geq\lambda_M(\mathbf{R}_{H_1})$, where $\lambda_m(\mathbf{R}_{H_0})=N\lambda_m(\hat{\mathbf{Q}}_{\mathbf{y},H_0})$ and $\lambda_m(\mathbf{R}_{H_1})=N\lambda_m(\hat{\mathbf{Q}}_{\mathbf{y},H_1})$.
Since $\mathbf{R}_{H_0}$ is a complex Gaussian Wishart matrix, its largest eigenvalue should follow Tracy-Widom distribution of order 2 \cite{chiani2014distribution}
\begin{equation}
\frac{\lambda_{\rm{max}}(\mathbf{R}_{H_0})-\mu}{\nu}\xrightarrow{d} \mathcal{TW}_2,
\end{equation}
where
\begin{subequations}
\begin{align}
&\mu=(\sqrt{M}+\sqrt{N})^2,\\
&\nu=\sqrt{\mu}\left(\frac{1}{\sqrt{M}}+\frac{1}{\sqrt{N}}\right)^{1/3},
\end{align}
\end{subequations}
are center and scaling parameters.
Then the cumulative distribution function (CDF) of $\mathcal{TW}_2$ is defined as
\begin{equation}
F_2(x)={\rm{exp}}\left\{-\int_x^\infty(a-x)q^2(a)da\right\},\label{tw2_cdf}
\end{equation}
where $q(a)$ is the solution of function
\begin{equation}
q''(a)=aq(a)+2q^3(a).\label{q_function}
\end{equation}

In addition, for the Wishart matrix $\mathbf{R}_{H_0}$, if $\lim\limits_{N\to+\infty}\frac{M}{N}=z~(z\in[0,1])$, its maximum and minimum eigenvalues can be approximated as $(\sqrt{N}+\sqrt{M})^2$ and $(\sqrt{N}-\sqrt{M})^2$ respectively. Next we will present several high-performance signal detectors based on the knowledge given earlier.

\subsection{Proposed SR-MME Detector}
The SR-MME detector is defined as square root of maximum eigenvalue times minimum eigenvalue, and is given by
\begin{equation}
\sqrt{\lambda_{\rm{max}}(\hat{\mathbf{Q}}_{\mathbf{y}})\lambda_{\rm{min}}(\hat{\mathbf{Q}}_{\mathbf{y}})}\mathop{\gtrless}\limits_{H_0}^{H_1}\gamma_1,
\end{equation}
where $\lambda_{\rm{max}}(\hat{\mathbf{Q}}_{\mathbf{y}})$, $\lambda_{\rm{min}}(\hat{\mathbf{Q}}_{\mathbf{y}})$ are maximum and minimum eigenvalues, respectively, of sample covariance matrix $\hat{\mathbf{Q}}_{\mathbf{y}}$, $\gamma_1$ denotes the judgment threshold.

At the end of judgment, there will be four possible results: true positive (TP), false positive (FP), true negative (TN), false negative (FN). In our work, only TP and FP are concerned, where the probability of FP is also called false alarm (FA) probability.
Therefore, $P_{FA}$ of SR-MME detector is defined as
\begin{equation}
\begin{aligned}
P_{FA}&=P\left(\sqrt{\lambda_{\rm{max}}(\hat{\mathbf{Q}}_{\mathbf{y},H_0})\lambda_{\rm{min}}(\hat{\mathbf{Q}}_{\mathbf{y},H_0})}>\gamma_1\right)\\
&=P\left(\lambda_{\rm{max}}(\mathbf{R}_{H_0})>\frac{(N\gamma_1)^2}{\lambda_{\rm{min}}(\mathbf{R}_{H_0})}\right)\\
&=P\left(\frac{\lambda_{\rm{max}}(\mathbf{R}_{H_0})-\mu}{\nu}>\frac{\left(\frac{N\gamma_1}{\sqrt{N}-\sqrt{M}}\right)^2-\mu}{\nu}\right)\\
&=1-F_2\left(\frac{\left(\frac{N\gamma_1}{\sqrt{N}-\sqrt{M}}\right)^2-\mu}{\nu}\right),
\end{aligned}
\end{equation}
then the threshold can be derived as
\begin{equation}
\gamma_1=\frac{\sqrt{N}-\sqrt{M}}{N}\sqrt{\nu F_2^{-1}(1-P_{FA})+\mu}.
\end{equation}

When the signal exists, sample covariance matrix (\ref{R_H1}) is no longer a wishart matrix. As shown in \cite{zeng2009eigenvalue}, it's maximum and minimum eigenvalues can be approximated as
\begin{subequations}
\begin{align}
\lambda_{\textrm{max}}(\mathbf{R}_{H_1})&=N\rho_1+\lambda_{\textrm{max}}(\mathbf{R}_{H_0}),\\
\lambda_{\textrm{min}}(\mathbf{R}_{H_1})&=N\rho_M+\sqrt{N}(\sqrt{N}-\sqrt{M}),
\end{align}
\end{subequations}
The detection probability, is given by
\begin{equation}
\begin{aligned}
&P_D=P\left(\sqrt{\lambda_{\rm{max}}(\hat{\mathbf{Q}}_{\mathbf{y},H_1})\lambda_{\rm{min}}(\hat{\mathbf{Q}}_{\mathbf{y},H_1})}>\gamma_1\right)\\
&=P\left(\lambda_{\rm{max}}(\mathbf{R}_{H_1})>\frac{(N\gamma_1)^2}{\lambda_{\rm{min}}(\mathbf{R}_{H_1})}\right)\\
&=P\left(\frac{\lambda_{\rm{max}}(\mathbf{R}_{H_0})-\mu}{\nu}>\frac{\frac{(N\gamma_1)^2}{N\rho_M+N-\sqrt{MN}}-N\rho_1-\mu}{\nu}\right)\\
&=1-F_2\left(\frac{\frac{(N\gamma_1)^2}{N\rho_M+N-\sqrt{MN}}-\rho_1-\mu}{\nu}\right).
\end{aligned}
\end{equation}
\subsection{Proposed GM Detector}
The geometric mean (GM) detector is defined as
\begin{equation}
\sqrt[M]{\prod_{m=1}^M\lambda_{m}(\hat{\mathbf{Q}}_{\mathbf{y}})}\mathop{\gtrless}\limits_{H_0}^{H_1}\gamma_2,
\end{equation}
where $\lambda_{m}(\hat{\mathbf{Q}}_{\mathbf{y}})$ is the eigenvalue of sample covariance matrix and $\gamma_2$ represents the judgment threshold of this detector. Similar to SR-MME detector, the false alarm probability of GM detector is given by
\begin{equation}
\begin{aligned}
&P_{FA}=P\left(\sqrt[M]{\prod_{m=1}^M\lambda_{m}(\hat{\mathbf{Q}}_{\mathbf{y},H_0})}>\gamma_2\right)\\
&=P\left(\lambda_{\rm{max}}(\mathbf{R}_{H_0})>\gamma_2^M\frac{\lambda_{\rm{max}}(\mathbf{R}_{H_0})}{{\rm{det}}(\hat{\mathbf{Q}}_{\mathbf{y},H_0})}\right)\\
&=P\left(\frac{\lambda_{\rm{max}}(\mathbf{R}_{H_0})-\mu}{\nu}>\frac{\gamma_2^M\frac{(\sqrt{N}+\sqrt{M})^2}{{\rm{det}}(\hat{\mathbf{Q}}_{\mathbf{y},H_0})}-\mu}{\nu}\right)\\
&=1-F_2\left(\frac{\gamma_2^M\frac{(\sqrt{N}+\sqrt{M})^2}{{\rm{det}}(\hat{\mathbf{Q}}_{\mathbf{y},H_0})}-\mu}{\nu}\right),
\end{aligned}
\end{equation}
and threshold is
\begin{equation}
\gamma_2=\sqrt[M]{\frac{\left(\nu F_2^{-1}(1-P_{FA})+\mu\right){\rm{det}}(\hat{\mathbf{Q}}_{\mathbf{y},H_0})}{(\sqrt{N}+\sqrt{M})^2}}.
\end{equation}

Finally, the detection probability of GM detector can be expressed by
\begin{equation}
\begin{aligned}
&P_D=P\left(\sqrt[M]{\prod_{m=1}^M\lambda_{m}(\hat{\mathbf{Q}}_{\mathbf{y},H_1})}>\gamma_2\right)\\
&=P\left(\lambda_{\rm{max}}(\mathbf{R}_{H_0})>\gamma_2^M\frac{\lambda_{\rm{max}}(\mathbf{R}_{H_0})}{{\rm{det}}(\hat{\mathbf{Q}}_{\mathbf{y},H_1})}\right)\\
&=1-F_2\left(\frac{\gamma_2^M\frac{(\sqrt{N}+\sqrt{M})^2}{{\rm{det}}(\hat{\mathbf{Q}}_{\mathbf{y},H_1})}-\mu}{\nu}\right).
\end{aligned}
\end{equation}

\section{Proposed Classifiers for Inferring The Number of Passive Emitters}\label{signal classifier}
Since the detectors proposed in Section \ref{signal detectors} are designed for detecting whether the signals received by base station are from UAV emitters or noise only. If the UAVs are present, we need to further determine their number. Therefore, a multi-layer neural network (ML-NN) classifier is given in the following. Support vector machine (SVM) classifier and naive bayes classifier (NBC) are also discussed as benchmarks.
\subsection{Feature Selection and Extraction}
As can be seen in Fig.\ref{system flowchart}, after the sampling of the received signal, taking eigenvalue decomposition on the sample covariance matrix $\hat{\mathbf{Q}}_{\mathbf{y}}$, we can get eigenvalues $\hat{\lambda}_1\geq\hat{\lambda}_2\geq\cdots\geq\hat{\lambda}_M$. Although the sample covariance matrix is only an approximation of actual received signal covariance matrix, its eigenvalues also approximately satisfy (\ref{eigenvalue_character}) if the sample number $N$ is large enough, i.e. the maximum $K$ eigenvalues belong to signal subspace. Therefore, this character can be used to determine the number of signal emitters. Firstly, the following features of $\{\hat{\lambda}_m\}_{m=1}^M$ are selected to construct the feature space of received signal $\mathbf{Y}$, where
\begin{equation}
\left\{
\begin{aligned}
&\hat{\lambda}_{\textrm{max}},~~\hat{\lambda}_{\textrm{min}}\\
&\bar{\lambda}=\frac{1}{M}\sum_{m=1}^M\hat{\lambda}_m,~~\tilde{\lambda}=\left(\prod_{m=1}^M\hat{\lambda}_m\right)^{1/M}\\
&\sigma_{\hat{\lambda}}=\sqrt{\frac{\sum_{m=1}^M(\hat{\lambda}_m-\bar{\lambda})^2}{M}},
\end{aligned}
\right.
\end{equation}
As the number of emitters grows, the features also increase. In order to enlarge the discrimination between the different signals, we perform log operations on them. Then, the feature vector of any received signal is given by
\begin{equation}
\mathbf{x}=\left(\log(\hat{\lambda}_{\textrm{max}}),\log(\hat{\lambda}_{\textrm{min}}),\log(\bar{\lambda}),\log(\tilde{\lambda}),\log(\sigma_{\hat{\lambda}})\right).\label{feature_vector}
\end{equation}

Since the signal received by the base station is derived from different emitters, and it is a typical multiclass problem, machine learning-based methods are very suitable. Assuming there are most $K$ emitters in the coverage area of base station, we can obtain a $K$-elements classifier based on the existing training data, and then substitute the signal to be detected into this classifier for classification. Then we will introduce several high performance classification algorithms.

\begin{figure}[ht]
  \centering
  \includegraphics[width=0.48\textwidth]{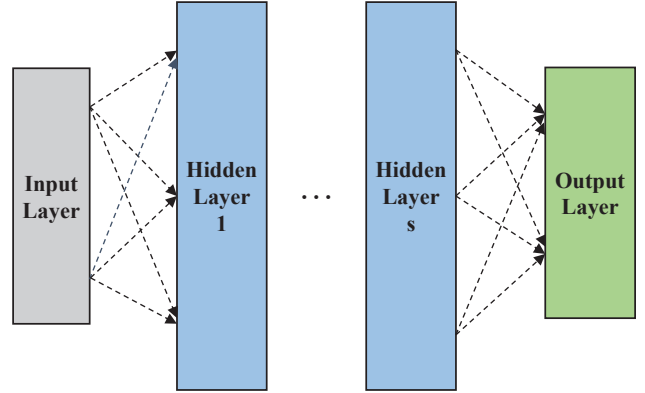}\\
  \caption{Multi-layer neural network.\label{Neural Network}}
\end{figure}
\subsection{Proposed Multi-layer Neural Network Classifier}
Given a set of received signals for training, such as $\mathbf{X}=\{(\mathbf{x}_i,\mathbf{g}_i)\}_{i=1,2,\cdots}$, where $\mathbf{g}_i=[g_{i,1},\cdots,g_{i,k},\cdots,g_{i,K}]$ is corresponding output vector.
It is a unit vector, if signal $i$ belongs to class $k$, $g_{i,k}=1$.
As is shown in Fig.\ref{Neural Network}, the input of this neural network is feature vector defined in (\ref{feature_vector}), the input layer is constructed of 5 neurons. Since there are most $K$ emitters in the coverage area of base station, the number of neurons in output layer is also $K$ and the outputs of these neurons are denoted by $\{\hat{g}_1,\hat{g}_2,\cdots,\hat{g}_K\}$. Assuming there are total $s$ hidden layers in this network, these hidden layers contain $q_1,q_2,\cdots,q_s$ neurons respectively. Therefore, referring to \cite{hagan1997neural}, the input that received by the $j_1$th neuron of hidden layer 1 can be represented as
\begin{equation}
\alpha_{1,j_1}=\sum_{h=1}^5v_{h,j_1}\mathbf{x}(h),
\end{equation}
where $v_{h,j_1}$ is the connection coefficient between the $h$th neuron of input layer and the $j_1$th neuron of hidden layer 1. Then, the output of this neuron is given by
\begin{equation}
z_{j_1}^1=f(\alpha_{1,j_1}-\delta_{1,j_1}),
\end{equation}
where $\delta_{1,j_1}$ denotes threshold of the $j$th neuron of hidden layer 1. $f(\cdot)$ is the activation function, and usually sigmoid function is adopted, which can be defined as
\begin{equation}
{\textrm{sigmoid}}(x)=\frac{1}{1+e^{-x}}.
\end{equation}

We can deduce input and output of the rest hidden layers from hidden layer 1, and the output from the $j_s$th neuron of hidden layer $s$ is given as
\begin{equation}
\begin{aligned}
z_{j_s}^s&=f(\alpha_{s,j_s}-\delta_{s,j_s})\\
&=f\left(\sum_{j_{s-1}=1}^{q_{s-1}}u_{j_{s-1},j_{s}}z_{j_{s-1}}^{s-1}-\delta_{s,j_s}\right),
\end{aligned}
\end{equation}
where $u_{j_{s-1},j_{s}}$ represents the connection coefficient between the $j_{s-1}$th neuron of hidden layer $s-1$ and the $j_s$th neuron of hidden layer $s$. Since output of the last hidden layer is transmitted to output layer, the final output of this network is
\begin{equation}
\hat{g}_k=f(\beta_k-\varepsilon_k)=f\left(\sum_{j_s=1}^{q_s}w_{j_s,k}z_{j_s}^s-\varepsilon_k\right),
\end{equation}
where $w_{j_s,k}$ is the connection coefficient between hidden layer $s$ and output layer, and $\varepsilon_k$ is threshold of the $k$th neuron of output layer.

When the input signal is $\mathbf{x}_1$, the ideal output is $\mathbf{g}_i$. However, the actual output of this neural network is $\hat{\mathbf{g}}_i=[\hat{g}_{i,1},\cdots,\hat{g}_{i,k},\cdots,\hat{g}_{i,K}]$, then the mean squared error (MSE) between ideal output and actual output is derived as
\begin{equation}
E_i=\frac{1}{K}\sum_{k=1}^K\left(\hat{g}_{i,k}-g_{i,k}\right)^2,
\end{equation}
Based on the classification error, we can update all the $(5q_1+\sum_{t=1}^{s-1}q_tq_{t+1}+q_sK)$ connection coefficients and $(\sum_{t=1}^sq_t+K)$ thresholds of this neural network. Taking the $j_s$th neuron of hidden layer $s$ as an example, we can get
\begin{subequations}
\begin{align}
w_{j_s,k}^{l+1}&=w_{j_s,k}^l+\Delta w_{j_s,k}^l,\\
\delta_{s,j_s}^{l+1}&=\delta_{s,j_s}^l+\Delta\delta_{s,j_s}^l,
\end{align}
\end{subequations}
where $l$ represents number of iterations. According to the gradient descent method, the update terms are defined as
\begin{equation}
\begin{aligned}
\Delta w_{j_s,k}^l&=-\eta\frac{\partial E_i}{\partial w_{j_s,k}^l}\\
&=-\eta\frac{\partial E_i}{\partial\hat{g}_{i,k}}\cdot\frac{\partial\hat{g}_{i,k}}{\partial\beta_{k}}\cdot\frac{\partial\beta_{k}}{\partial w_{j_s,k}^l}\\
&=-\frac{2\eta}{K}z_{j_s}^s\cdot G_{i,k},
\end{aligned}
\end{equation}
and
\begin{equation}
\begin{aligned}
\Delta\delta_{s,j_s}^l&=-\eta\frac{\partial E_i}{\partial \delta_{s,j_s}^l}\\
&=-\eta\sum_{k=1}^K\frac{\partial E_i}{\partial\hat{g}_{i,k}}\cdot\frac{\partial\hat{g}_{i,k}}{\partial\beta_{k}}\cdot\frac{\partial\beta_{k}}{\partial z_{j_s}^s}\cdot\frac{\partial z_{j_s}^s}{\partial\delta_{s,j_s}^l}\\
&=-\frac{2\eta}{K}z_{j_s}^s(1-z_{j_s}^s)\cdot\sum_{k=1}^K w_{j_s,k}^lG_{i,k},
\end{aligned}
\end{equation}
where $\eta$ is learning rate and
\begin{equation}
G_{i,k}=\hat{g}_{i,k}(1-\hat{g}_{i,k})(\hat{g}_{i,k}-g_{i,k}).
\end{equation}

All the parameters in the neural network are updated in each iteration until the parameters change less than a certain threshold or a certain number of iterations is reached. Therefore, the final classification result for signal $i$ is given by
\begin{equation}
C_i=\arg\max_k~\hat{g}_{i,k}^L,
\end{equation}
where $C_i\in\{1,2,\cdots,K\}$.

\subsection{Support Vector Machine Classifier}
Since determining the number of signal sources is a $K$-elements classification problem, it can be decomposed into $K(K-1)/2$ binary classification problems and each of these binary classification problems can be solved by support vector machine (SVM) method.
Given a training sample set $D=\{(\mathbf{x}_1,g_1),(\mathbf{x}_2,g_2),\cdots,(\mathbf{x}_s,g_s)\}$, where $g_i=\{-1,+1\}$. $g_i=-1$ denotes that signal $i$ belongs to class 1 and $g_i=+1$ denotes that this signal belongs to class 2. The separable hyperplane for sample space can be expressed by
\begin{equation}
\mathbf{w}^T\mathbf{x}+b=0,
\end{equation}
where $\mathbf{w}$ is normal vector which determines the direction of this hyperplane, and $b$ denotes the bias which is defined as the distance from hyperplane to original point. Therefore, the separable hyperplane can be denoted as $(\mathbf{w},b)$.

Assuming the samples can be classified by hyperplane $(\mathbf{w},b)$ accurately, if $g_i=-1$ we can get $\mathbf{w}^T\mathbf{x}_i+b<0$, and if $g_i=+1$ we get $\mathbf{w}^T\mathbf{x}_i+b>0$. Then the following conditions should be satisfied
\begin{equation}
\left\{
\begin{aligned}
&\mathbf{w}^T\mathbf{x}_i+b\geq +1,~g_i=+1\\
&\mathbf{w}^T\mathbf{x}_i+b\leq -1,~g_i=-1,
\end{aligned}
\right.\label{svm_condition}
\end{equation}
the samples closest to the separable hyperplane make the equalities in (\ref{svm_condition}) hold, and they are support vectors. The sum of the distance from the two heterologous support vectors to the hyperplane is called margin, and it is defined as $\delta=\frac{2}{\|\mathbf{w}\|}$. For maximizing the margin of separable hyperplane, the optimization problem can be designed as
\begin{subequations}
\begin{align}
\min_{\mathbf{w},b}~~&\frac{1}{2}\|\mathbf{w}\|^2\\
\textrm{s.t.}~~&g_i(\mathbf{w}^T\mathbf{x}_i+b)\geq 1.
\end{align}\label{svm_optimization}
\end{subequations}

Actually, the training samples can hardly be linearly separated in the current sample space. Firstly, we map the samples to a higher dimensional feature space. The the model of separable hyperplane is modified as
\begin{equation}
f(\mathbf{x})=\mathbf{w}^T\phi(\mathbf{x})+b,
\end{equation}
Secondly, to avoid overfitting, we introduce the concept of soft margin. This concept allows SVM to make errors in the classification of some samples, i.e., these samples can not satisfy constraint $g_i(\mathbf{w}^T\phi(\mathbf{x}_i)+b)\geq 1$. Consequently, the optimization problem (\ref{svm_optimization}) is transformed to maximize the margin while minimizing the classification error
\begin{subequations}
\begin{align}
\min_{\mathbf{w},b,\xi_i}~~&\frac{1}{2}\|\mathbf{w}\|^2+C\sum_{i=1}^s\xi_i,\\
\textrm{s.t.}~~&g_i(\mathbf{w}^T\phi(\mathbf{x}_i)+b)\geq 1-\xi_i,\\
&\xi_i\geq 0.
\end{align}\label{svm_soft_margin}
\end{subequations}
where $C>0$ is regularization constant, $\xi_i\geq 0$ is a slack variable and $\xi_i\geq 1$ means sample $\mathbf{x}_i$ is misclassified.

Obviously, (\ref{svm_soft_margin}) is a quadratic programming (QP) problem, and it can be solved by Lagrangian multiplier method. Therefore, the Lagrangian of (\ref{svm_soft_margin}) is given by
\begin{equation}
\begin{aligned}
L(\mathbf{w},b,\bm{\xi},\bm{\alpha},\bm{\beta})&=\frac{1}{2}\|\mathbf{w}\|^2+C\sum_{i=1}^s\xi_i-\sum_{i=1}^s\beta_i\xi_i\\
&+\sum_{i=1}^s\alpha_i\left[1-\xi_i-g_i(\mathbf{w}^T\phi(\mathbf{x}_i)+b)\right],\label{lagrangian}
\end{aligned}
\end{equation}
where $\alpha_i\geq 0$ and $\beta_i\geq 0$ are Lagrangian multipliers. Computing the partial derivatives of $\mathbf{w},b,\xi_i$, we can get
\begin{subequations}
\begin{align}
&\mathbf{w}=\sum_{i=1}^s\alpha_ig_i\phi(\mathbf{x}_i),\\
&\sum_{i=1}^s\alpha_ig_i=0,\label{alpha_constraint}\\
&C=\alpha_i+\beta_i,
\end{align}
\end{subequations}
taking them into equation (\ref{lagrangian}), the dual problem of (\ref{svm_soft_margin}) is derived as
\begin{subequations}
\begin{align}
\max_{\alpha_i}~~&\sum_{i=1}^s\alpha_i-\frac{1}{2}\sum_{i=1}^s\sum_{j=1}^s\alpha_i\alpha_jg_ig_j\kappa(\mathbf{x}_i,\mathbf{x}_j),\\
\textrm{s.t.}~~&(\ref{alpha_constraint}),\\
&0\leq\alpha_i\leq C,
\end{align}
\end{subequations}
where $\kappa(\mathbf{x}_i,\mathbf{x}_j)=\phi(\mathbf{x}_i)^T\phi(\mathbf{x}_j)$ is the kernel function.

Since (\ref{svm_soft_margin}) contains inequality constraint, the above optimization procedure must satisfy the KKT conditions
\begin{equation}
\left\{
\begin{aligned}
&\alpha_i\geq0,~\beta_i\geq0\\
&g_if(\mathbf{x}_i)-1+\xi_i\geq0\\
&\alpha_i(g_if(\mathbf{x}_i)-1+\xi_i)=0\\
&\xi_i\geq0,~\beta_i\xi_i=0.
\end{aligned}
\right.
\end{equation}
\subsection{Naive Bayes Classifier}
As given in (\ref{feature_vector}), three features of $i$th signal are considered in our problem. We assume that the 5 features are independent of each other, then according to bayes theorem, the probability that the $i$th signal belongs to a certain class is
\begin{equation}
P(c_k|\mathbf{x}_i)=\frac{P(c_k)P(\mathbf{x}_i|c_k)}{P(\mathbf{x}_i)}=\frac{P(c_k)P(\mathbf{x}_i|c_k)}{\sum_{k=1}^K P(\mathbf{x}_i|c_k)P(c_k)},
\end{equation}
where $c_k,~k\in D=\{1,2,\cdots,K\}$ is the label for classification. Therefore, the NBC for our problem can be verified as
\begin{equation}
h(\mathbf{x}_i)=\arg\max_{k\in D}P(c_k)P(\mathbf{x}_i|c_k).\label{NBC}
\end{equation}

The training process is based on the training set to estimate the class prior probability $P(c_k)$ and conditional probability $P(\mathbf{x}_i|c_k)$. Since the features in (\ref{feature_vector}) are continuous, we can suppose $P(\mathbf{x}_i|c_k)\sim \mathcal{N}(\mu_k,\Sigma_k)$, where $\mu_k$ and $\Sigma_k$ are mean and covariance matrix of feature vectors for all train samples that belong to class $k$. Therefore, the conditional probability can be represented by its PDF as
\begin{equation}
P(\mathbf{x}_i|c_k)=\frac{1}{(\sqrt{2\pi})^5|\Sigma|^{1/2}}e^{-\frac{1}{2}(\mathbf{x}_i-\mu_k)^T\Sigma_k^{-1}(\mathbf{x}_i-\mu_k)},
\end{equation}
then, we can compute logarithm of (\ref{NBC}). Finally, the NBC can be transformed as
\begin{equation}
\begin{aligned}
h(\mathbf{x}_i)=\arg\max_{k\in D} &\ln\left(P(c_k)P(\mathbf{x}_i|c_k)\right)\\
=\arg\max_{k\in D}&\left(\ln P(c_k)-\frac{5}{2}\ln2\pi-\frac{1}{2}\ln|\Sigma_k|\right.\\
&\left.-\frac{1}{2}(\mathbf{x}_i-\mu_k)^T\Sigma_k^{-1}(\mathbf{x}_i-\mu_k)\right).
\end{aligned}
\end{equation}

\section{Simulation Results}\label{simulation}
In this section, representative  simulation results are given to show the high performance of signal detectors and classifiers proposed in this paper. Next, we will compare the two proposed signal detectors with existing detectors.

\begin{table*}[htb]
\centering
\caption{\scshape{Numerical Table For The Tracy-Widom Distribution of Order 2}}\label{table1}
\scalebox{1.05}
{\begin{tabular}{|c|c|c|c|c|c|c|c|c|c|}
  \hline
  $t$ & -3.70 & -2.90 & -1.80 & -0.60 & -0.23 & 0.49 & 1.32 & 2.06 & 2.68\\
  \hline $F_2(t)$ & 0.01 & 0.1 & 0.5 & 0.9 & 0.95 & 0.99 & 0.999 & 0.9999 & 0.99999 \\
  \hline
\end{tabular}}
\end{table*}

\subsection{Signal Detectors}
Firstly, it is assumed that there are 3 UAV emitters in the coverage area of base station, i.e. $K=3$ and the signals used in this simulation are randomly generated signals. After sampling the received signal, we can obtain the sample covariance matrix. The largest eigenvalue of noise-only sample covariance matrix ($\mathbf{R}_{H_0}$) follows Tracy-Widom distribution of order 2, so that we want to use its statistical properties to derive $P_{FA}$, $P_D$ and $\gamma$ of signal detectors. But (\ref{q_function}) is difficult to evaluate, since we cannot obtain the CDF of $\mathcal{TW}_2$. Fortunately, M. Pr{\"a}hofer and H. Spohn fitted this function and gave tables for CDF of Tracy-Widom distribution in \cite{prahofer2004exact}. We may select a part of the values and put them in Table \ref{table1}. To highlight the advantages of our proposed signal detectors, we also introduce two existing detectors for comparison. The two detectors, M-MME and MME \cite{zeng2009eigenvalue}, are defined as
\begin{subequations}
\begin{align}
\textbf{M-MME}:~&\frac{\lambda_{\textrm{max}}(\hat{\mathbf{Q}}_{\mathbf{y}})+\lambda_{\textrm{min}}(\hat{\mathbf{Q}}_{\mathbf{y}})}{2}\mathop{\gtrless}\limits_{H_0}^{H_1}\gamma_4,\\
\textbf{MME}:~&\frac{\lambda_{\textrm{max}}(\hat{\mathbf{Q}}_{\mathbf{y}})}{\lambda_{\textrm{min}}(\hat{\mathbf{Q}}_{\mathbf{y}})}\mathop{\gtrless}\limits_{H_0}^{H_1}\gamma_3.
\end{align}
\end{subequations}

\begin{figure}[htb]
  \centering
    \includegraphics[width=0.48\textwidth]{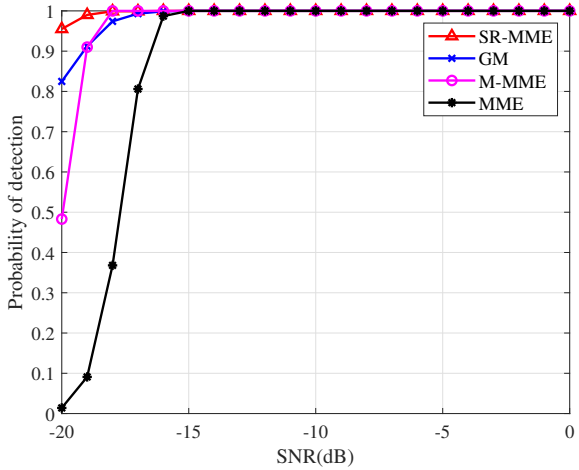}\\
  \caption{Probability of detection versus SNR, $P_{FA}=10^{-4}$, $N=200$.}\label{pd_snr}
\end{figure}

As can be seen in Fig.\ref{pd_snr}, the relationship between SNR and probability of detection is plotted, where probability of false alarm $P_{FA}=10^{-4}$, number of receive antennas $M=64$, and number of samples $N=200$. Among these four detectors, SR-MME has the best performance across all SNR values, and its detection probability of signal sources reaches $95\%$ even with poor SNR environment. When SNR=-20dB, the detection probability of SR-MME increases more than 90\% compared with MME and also exceeds that of MME nearly 50\%. For the GM detector, its detection probability is slightly less than SR-MME at low SNR situation, but it still has a great improvement compared to the other two detectors.

\begin{figure}[htb]
  \centering
    \includegraphics[width=0.48\textwidth]{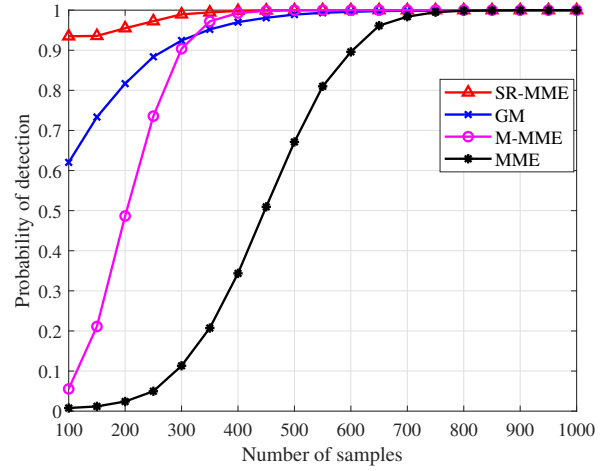}\\
  \caption{Probability of detection versus number of samples, SNR=-20dB, $P_{FA}=10^{-4}$.}\label{pd_N}
\end{figure}

Fig.\ref{pd_N} presents the detection probability of these four signal detectors with the number of samples, where $M=64$, $P_{FA}=10^{-4}$ and SNR=-20dB. The overall trend of the curves in this figure is similar to Fig.\ref{pd_snr}, with SR-MME still the best performing of these four signal detectors and achieving a detection probability of at least 93\%. The detection performance of GM detector also improves as the number of samples increases, especially when $N$ ranges between 100 and 200. GM has a significant improvement compared with M-MME and MME. Therefore, the robust performance of SR-MME and GM at lower number of samples can help us save lots of time and spatial resources, and not at the cost of a loss of detection performance.

\begin{figure}[htb]
  \centering
    \includegraphics[width=0.48\textwidth]{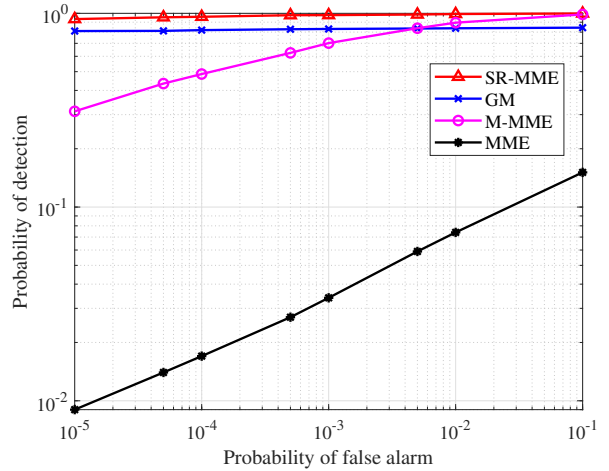}\\
  \caption{ROC curve, SNR=-20dB, $N=200$.}\label{roc}
\end{figure}

Fig.\ref{roc} shows the most commonly used indicator in the field of threshold detection, the Receiver Operating Characteristic (ROC) curve. It evaluates a detector comprehensively in terms of both detection probability and false alarm probability. The parameters involved in this simulation are $M=64$, $N=200$ and SNR=-20dB. The  ROC curve of SR-MME is above the other three curves, so it is the best detector for the overall performance. Correspondingly, the MME has the worst performance. For GM and M-MME, due to a cross-over of their ROC curves, area under ROC curve (AUC) is introduced for comparing their performance. Since the axes in this figure employ scientific counting, after converting it to ordinary coordinates, the AUC value of M-MME is larger than GM. From this perspective, M-MME performs better than GM. But in practice, we would prefer a relatively low false alarm probability, so GM will be more useful for it can guarantee a low false alarm probability while maintaining a high detection probability.

\begin{table*}[htb]
\centering
\caption{\scshape{Average Training Duration of Different Classifiers}}\label{table2}
\scalebox{1.05}
{\begin{tabular}{|c|c|c|c|c|c|c|}
  \hline
  \multirow{2}*{\textbf{Classifiers}}& \multicolumn{6}{|c|}{\textbf{Number of Training Samples}}\\\cline{2-7}
  ~ & 10 & 20 & 30 & 40 & 50 & 100 \\\hline
  4-layer Neural Network & 0.734149 & 0.809213 & 0.936686 & 1.038361 & 1.133686 & 1.660306 \\\hline
  3-layer Neural Network & 0.629034 & 0.705787 & 0.799842 & 0.875255 & 0.949917 & 1.356083 \\\hline
  SVM & 0.221015 & 0.333413 & 0.520857 & 0.753500 & 1.007692 & 3.077889 \\\hline
  NBC & 0.090488 & 0.092070 & 0.093222 & 0.094849 & 0.095326 & 0.113129 \\\hline
\end{tabular}}
\end{table*}
\subsection{Signal Classifiers}
After the presence of the emitters is determined by the signal detectors, we need to further determine the number of emitters, and this process is called signal classification. According to the three machine learning-based signal classifiers, the first step is to design an appropriate training set. As mentioned in Section \ref{signal classifier}, the feature vector of received signals is given by (\ref{feature_vector}), so the training set is defined as
\begin{equation}
\left\{\mathbf{X}_1,\cdots,\mathbf{X}_k,\cdots,\mathbf{X}_K\right\},
\end{equation}
where
\begin{equation}
\mathbf{X}_k=\left\{\left(\mathbf{x}_{k,1},k\right),\left(\mathbf{x}_{k,2},k\right),\cdots,\left(\mathbf{x}_{k,i},k\right),\cdots\right\},
\end{equation}
and $K\in \{1,2,3\}$. For the training of ML-NN, the epoch size is 400, learning rate is set as 0.01. The input layer and output layer have 5 neurons and 3 neurons respectively, the hidden layer size of 3-layer NN is 10,  4-layer NN has 2 hidden layers and their size are 7 and 5.

In order to compare the complexity of the ML-based methods mentioned in our work, Table \ref{table2} gives the training duration of each classifier at different amounts of training data. The neural network takes more training time as the number of training samples is small. When the amount of training data reaches 50, the average training duration of SVM exceeds the 3-layer neural network. Different from other classifiers, the change in the number of training samples has less impact on NBC.

\begin{figure}[htb]
  \centering
  \includegraphics[width=0.48\textwidth]{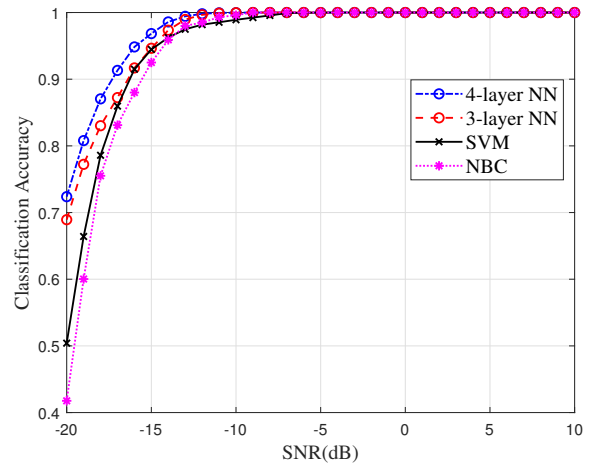}\\
  \caption{Classification accuracy versus SNR, $M=64$.}\label{snr_accuracy}
\end{figure}

Fig.\ref{snr_accuracy} plots the relationship between the classification accuracy of the four classifiers and SNR, where $M=64$, $N=200$ and the number of training samples is 10. It can be seen that neural networks have excellent classification performance at low SNR situation. Even in the extreme case of -20dB, 3-layer neural network still achieves nearly 70\% classification accuracy, and 4-layer neural network can improve further. After simulation attempts, 4-layer neural network is optimal for our classification problem. Since neural networks have strong learning ability, the deeper networks can instead cause overfitting, and result in the decrease of classification accuracy. The performance of SVM can be close to neural network at -15dB, and the accuracy of NBC is lower than SVM.

By observing the curves of the signal detectors and the signal classifiers about SNR in Fig.\ref{pd_snr} and Fig.\ref{snr_accuracy}, we can find when SNR=-20dB and $P_{FA}=10^{-4}$, the $P_D$ of SR-MME can achieve 95\%. Since $P_{FA}+P_{AN}=1$, SR-MME almost separates all the noise while ensuring a high signal detection rate. However, for the optimal neural network-based signal classifier, its classification accuracy at SNR= -20dB is also only about 70\%, that is, if the noise is directly added to the classification process, nearly 30\% of the noise will be misclassified as signals. Therefore, we believe that adding the step of signal detection is necessary. Moreover, the time required to perform one signal detection was approximately 0.04s, and the training duration required for the 4-layer neural network after adding noise is also increased to about 1.02s when number of training sample is 10. Therefore, using the signal detectors can also save the time cost.

\begin{figure}[htb]
  \centering
  \includegraphics[width=0.48\textwidth]{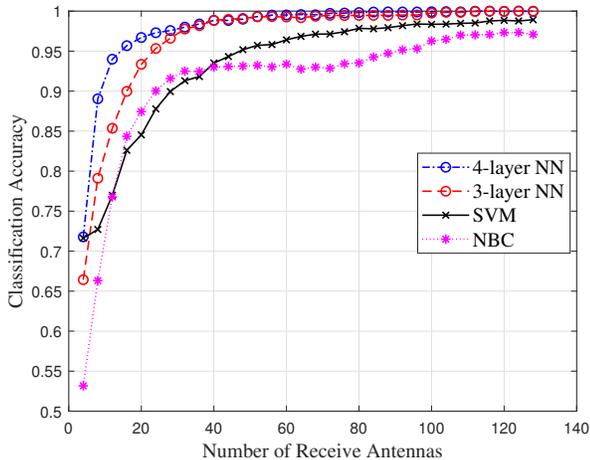}\\
  \caption{Classification accuracy versus number of receive antennas, SNR=-15dB.}\label{accuracy_antenna}
\end{figure}

In Fig.\ref{accuracy_antenna}, we show classification accuracy varying with the number of received antennas when SNR=-15dB, and other conditions are the same as Fig.\ref{snr_accuracy}. In general, array containing 64 antennas and more can be called massive array. Therefore, as can be seen in this figure, the classification accuracy of neural networks can approach nearly 100\% when a massive receive array is adopted. The performance of SVM and NBC is worse than neural network with a massive receive array.

\subsection{Analysis of Classic Classifiers}
AIC and MDL are two classic information theoretic criteria for model selection, which were proposed by Akaike \cite{akaike1974new} \cite{akaike1998information}, Schwartz \cite{schwarz1978estimating} and Rissanen \cite{rissanen1978modeling}. In Akaike's works, the AIC criterion is defined as
\begin{equation}
	\textrm{AIC}(m)=-2\log L_m^{(M-m)N}+2m(2M-m),
\end{equation}
where $m\in\{0,1,\cdots,M-1\}$ and
\begin{equation}
	L_m=\frac{\prod_{i=m+1}^M \hat{\lambda}_i^{1/(M-m)}}{\frac{1}{M-m}\sum_{i=m+1}^M \hat{\lambda}_i},
\end{equation}
the classification results of received signals are determined by AIC criterion as following
\begin{equation}
	\textrm{AIC}(C)=\min\left(\textrm{AIC}(0),\textrm{AIC}(1),\cdots,\textrm{AIC}(M-1)\right),
\end{equation}
where $C$ is the number of emitters.

Similarly, the definition of MDL criterion is given as
\begin{equation}
	\textrm{MDL}(m)=-2\log L_m^{(M-m)N}+\frac{1}{2}m(2M-m)\log N,
\end{equation}
MDL modified the bias term based on AIC, leading to the improvement classification performance. The classification result of MDL is
\begin{equation}
	\textrm{MDL}(C)=\min\left(\textrm{MDL}(0),\textrm{MDL}(1),\cdots,\textrm{MDL}(M-1)\right).
\end{equation}

\begin{figure}[htb]
  \centering
  \includegraphics[width=0.48\textwidth]{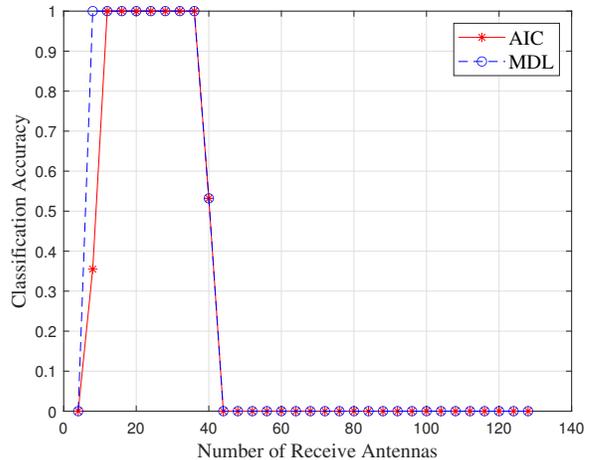}\\
  \caption{Classification accuracy versus number of receive antennas for AIC and MDL, SNR=0dB.}\label{accuracy_antenna_aic_mdl}
\end{figure}

Since the former papers only verified the work performance of AIC and MDL with small size receiving array, such as around 8 antennas. To find out whether these two methods can maintain good performance with massive receive array, we present a curve between their classification accuracy and the number of receive antennas. Unfortunately, as shown in Fig.\ref{accuracy_antenna_aic_mdl}, AIC and MDL can only achieve good performance when the number of receive antennas is between 8 and 36. Once the number of receive antennas exceeds 36, their classification accuracy drops sharply until the number of emitters is completely inaccessible at 44 antennas. By analyzing the definitions of AIC and MDL, since the number of receive antennas is equal to the number of possible classifications, the corresponding model complexity increases when the number of antennas increases. If the model is too complex, the values of AIC and MDL will increase, and resulting in overfitting. Thus, we can conclude that AIC and MDL are not applicable for scenarios using massive receive arrays. 

\begin{figure}[htb]
  \centering
  \includegraphics[width=0.48\textwidth]{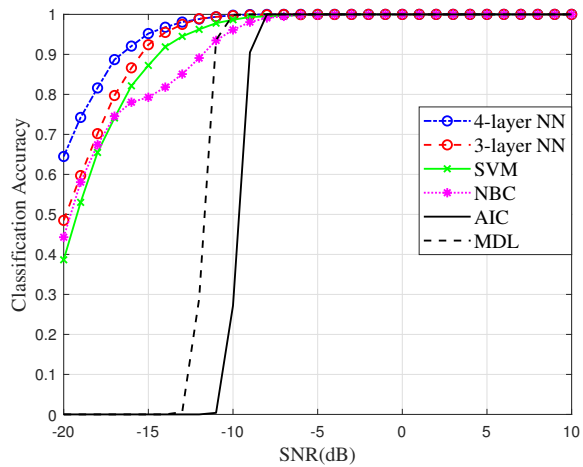}\\
  \caption{Classification accuracy versus SNR, $M=32$.}\label{accuracy_snr_32antennas}
\end{figure}

To compare the performance differences between traditional and machine learning-based methods, we plot the classification accuracy of these methods with SNR in Fig.\ref{accuracy_snr_32antennas}, where $M=32$. Although this is not in massive array scenario, the machine learning-based method still have higher classification accuracy than the AIC and MDL. Therefore, machine learning-based signal classifiers are robust and are applicable to a broader SNR range and array size.

\section{Conclusion}\label{conclusion}
In our work, a DOA preprocessing system containing signal detectors and ML-based signal classifiers has been proposed for inferring the number of UAV emitters in a massive MIMO system. We derived the theoretical thresholds and probability of detection for SR-MME and GM with the aid of random matrix theories. Simulation results showed that the proposed SR-MME and GM have much better performance than existing detectors like MME and M-MME, especially in the low SNR region and small number of samples situations. After determining the presence of emitters, we further inferred their number based on machine learning classifiers, by proposing ML-NN, SVM and NB. The classification accuracy of the proposed ML-NN is higher than that of other two classifiers. Finally, we also compared proposed methods with traditional methods AIC and MDL, and the proposed ML-NN was shown more applicable to scenarios with low SNR and massive MIMO receive arrays. In conclusion, we believe that the proposed system and method will be helpful for the future implementation of UAV massive MIMO communications.
\bibliographystyle{IEEEtran}
\bibliography{machine_learning}

\begin{thebibliography}{10}
\providecommand{\url}[1]{#1}
\csname url@samestyle\endcsname
\providecommand{\newblock}{\relax}
\providecommand{\bibinfo}[2]{#2}
\providecommand{\BIBentrySTDinterwordspacing}{\spaceskip=0pt\relax}
\providecommand{\BIBentryALTinterwordstretchfactor}{4}
\providecommand{\BIBentryALTinterwordspacing}{\spaceskip=\fontdimen2\font plus
\BIBentryALTinterwordstretchfactor\fontdimen3\font minus
  \fontdimen4\font\relax}
\providecommand{\BIBforeignlanguage}[2]{{%
\expandafter\ifx\csname l@#1\endcsname\relax
\typeout{** WARNING: IEEEtran.bst: No hyphenation pattern has been}%
\typeout{** loaded for the language `#1'. Using the pattern for}%
\typeout{** the default language instead.}%
\else
\language=\csname l@#1\endcsname
\fi
#2}}
\providecommand{\BIBdecl}{\relax}
\BIBdecl

\bibitem{zeng2016wireless}
Y.~Zeng, R.~Zhang, and T.~J. Lim, ``Wireless communications with unmanned
  aerial vehicles: Opportunities and challenges,'' \emph{IEEE Commun. Mag.},
  vol.~54, no.~5, pp. 36--42, 2016.

\bibitem{huang2021massive}
Y.~Huang, Q.~Wu, R.~Lu, X.~Peng, and R.~Zhang, ``Massive mimo for
  cellular-connected uav: Challenges and promising solutions,'' \emph{IEEE
  Commun. Mag.}, vol.~59, no.~2, pp. 84--90, 2021.

\bibitem{wang2014cellular}
C.-X. Wang, F.~Haider, X.~Gao, X.-H. You, Y.~Yang, D.~Yuan, H.~M. Aggoune,
  H.~Haas, S.~Fletcher, and E.~Hepsaydir, ``Cellular architecture and key
  technologies for 5g wireless communication networks,'' \emph{IEEE Commun.
  Mag.}, vol.~52, no.~2, pp. 122--130, 2014.

\bibitem{saad2019vision}
W.~Saad, M.~Bennis, and M.~Chen, ``A vision of 6g wireless systems:
  Applications, trends, technologies, and open research problems,'' \emph{IEEE
  Netw.}, vol.~34, no.~3, pp. 134--142, 2019.

\bibitem{zhang20196g}
Z.~Zhang, Y.~Xiao, Z.~Ma, M.~Xiao, Z.~Ding, X.~Lei, G.~K. Karagiannidis, and
  P.~Fan, ``6g wireless networks: Vision, requirements, architecture, and key
  technologies,'' \emph{IEEE Veh. Technol. Mag.}, vol.~14, no.~3, pp. 28--41,
  2019.

\bibitem{chandhar2019massive}
P.~Chandhar and E.~G. Larsson, ``Massive mimo for connectivity with drones:
  Case studies and future directions,'' \emph{IEEE Access}, vol.~7, pp.
  94\,676--94\,691, 2019.

\bibitem{harris2017performance}
P.~Harris, S.~Malkowsky, J.~Vieira, E.~Bengtsson, F.~Tufvesson, W.~B. Hasan,
  L.~Liu, M.~Beach, S.~Armour, and O.~Edfors, ``Performance characterization of
  a real-time massive mimo system with los mobile channels,'' \emph{IEEE J.
  Sel. Areas Commun.}, vol.~35, no.~6, pp. 1244--1253, 2017.

\bibitem{geraci2022will}
G.~Geraci, A.~Garcia-Rodriguez, M.~M. Azari, A.~Lozano, M.~Mezzavilla,
  S.~Chatzinotas, Y.~Chen, S.~Rangan, and M.~Di~Renzo, ``What will the future
  of uav cellular communications be? a flight from 5g to 6g,'' \emph{IEEE
  Commun. surveys Tuts.}, vol.~24, no.~3, pp. 1304--1335, 2022.

\bibitem{bai2022non}
L.~Bai, Z.~Huang, and X.~Cheng, ``A non-stationary model with time-space
  consistency for 6g massive mimo mmwave uav channels,'' \emph{IEEE Trans.
  Wireless Commun.}, 2022.

\bibitem{chandhar2017massive}
P.~Chandhar, D.~Danev, and E.~G. Larsson, ``Massive mimo for communications
  with drone swarms,'' \emph{IEEE Trans. Wireless Commun.}, vol.~17, no.~3, pp.
  1604--1629, 2017.

\bibitem{huang2015source}
L.~Huang, C.~Qian, H.~C. So, and J.~Fang, ``Source enumeration for large array
  using shrinkage-based detectors with small samples,'' \emph{IEEE Trans.
  Aerosp. Electron. Syst.}, vol.~51, no.~1, pp. 344--357, 2015.

\bibitem{krim1996two}
H.~Krim and M.~Viberg, ``Two decades of array signal processing research: the
  parametric approach,'' \emph{IEEE Signal Process. Mag.}, vol.~13, no.~4, pp.
  67--94, 1996.

\bibitem{bjornson2019massive}
E.~Bj{\"o}rnson, L.~Sanguinetti, H.~Wymeersch, J.~Hoydis, and T.~L. Marzetta,
  ``Massive mimo is a reality—what is next?: Five promising research
  directions for antenna arrays,'' \emph{Digital Signal Processing}, vol.~94,
  pp. 3--20, 2019.

\bibitem{akaike1974new}
H.~Akaike, ``A new look at the statistical model identification,'' \emph{IEEE
  Trans. Autom. Control}, vol.~19, no.~6, pp. 716--723, 1974.

\bibitem{schwarz1978estimating}
G.~Schwarz, ``Estimating the dimension of a model,'' \emph{The annals of
  statistics}, pp. 461--464, 1978.

\bibitem{rissanen1978modeling}
J.~Rissanen, ``Modeling by shortest data description,'' \emph{Automatica},
  vol.~14, no.~5, pp. 465--471, 1978.

\bibitem{stoica2004model}
P.~Stoica and Y.~Selen, ``Model-order selection: a review of information
  criterion rules,'' \emph{IEEE Signal Process. Mag.}, vol.~21, no.~4, pp.
  36--47, 2004.

\bibitem{lu2012generalized}
Z.~Lu and A.~M. Zoubir, ``Generalized bayesian information criterion for source
  enumeration in array processing,'' \emph{IEEE Trans. Signal Process.},
  vol.~61, no.~6, pp. 1470--1480, 2012.

\bibitem{lu2012flexible}
------, ``Flexible detection criterion for source enumeration in array
  processing,'' \emph{IEEE Trans. Signal Process.}, vol.~61, no.~6, pp.
  1303--1314, 2012.

\bibitem{williams1990using}
D.~B. Williams and D.~H. Johnson, ``Using the sphericity test for source
  detection with narrow-band passive arrays,'' \emph{IEEE Trans. Acoust.,
  Speech, Signal Process.}, vol.~38, no.~11, pp. 2008--2014, 1990.

\bibitem{brcich2002detection}
R.~F. Brcich, A.~M. Zoubir, and P.~Pelin, ``Detection of sources using
  bootstrap techniques,'' \emph{IEEE Trans. Signal Process.}, vol.~50, no.~2,
  pp. 206--215, 2002.

\bibitem{wax2021detection}
M.~Wax and A.~Adler, ``Detection of the number of signals by signal subspace
  matching,'' \emph{IEEE Trans. Signal Process.}, vol.~69, pp. 973--985, 2021.

\bibitem{cabric2004implementation}
D.~Cabric, S.~M. Mishra, and R.~W. Brodersen, ``Implementation issues in
  spectrum sensing for cognitive radios,'' in \emph{Conference Record of the
  Thirty-Eighth Asilomar Conference on Signals, Systems and Computers, 2004.},
  vol.~1.\hskip 1em plus 0.5em minus 0.4em\relax Ieee, 2004, pp. 772--776.

\bibitem{cabric2006spectrum}
D.~Cabric, A.~Tkachenko, and R.~W. Brodersen, ``Spectrum sensing measurements
  of pilot, energy, and collaborative detection,'' in \emph{Milcom 2006-2006
  IEEE military communications conference}.\hskip 1em plus 0.5em minus
  0.4em\relax IEEE, 2006, pp. 1--7.

\bibitem{gardner1991exploitation}
W.~A. Gardner, ``Exploitation of spectral redundancy in cyclostationary
  signals,'' \emph{IEEE Signal Process. Mag.}, vol.~8, no.~2, pp. 14--36, 1991.

\bibitem{zeng2009eigenvalue}
Y.~Zeng and Y.-C. Liang, ``Eigenvalue-based spectrum sensing algorithms for
  cognitive radio,'' \emph{IEEE Trans. Commun.}, vol.~57, no.~6, pp.
  1784--1793, 2009.

\bibitem{zhang2010multi}
R.~Zhang, T.~J. Lim, Y.-C. Liang, and Y.~Zeng, ``Multi-antenna based spectrum
  sensing for cognitive radios: A glrt approach,'' \emph{IEEE Trans. Commun.},
  vol.~58, no.~1, pp. 84--88, 2010.

\bibitem{liu2016optimal}
C.~Liu, H.~Li, J.~Wang, and M.~Jin, ``Optimal eigenvalue weighting detection
  for multi-antenna cognitive radio networks,'' \emph{IEEE Trans. Wireless
  Commun.}, vol.~16, no.~4, pp. 2083--2096, 2016.

\bibitem{anderson2004model}
D.~Anderson and K.~Burnham, ``Model selection and multi-model inference,''
  \emph{Second. NY: Springer-Verlag}, vol.~63, no. 2020, p.~10, 2004.

\bibitem{jiang2016machine}
C.~Jiang, H.~Zhang, Y.~Ren, Z.~Han, K.-C. Chen, and L.~Hanzo, ``Machine
  learning paradigms for next-generation wireless networks,'' \emph{IEEE
  Wireless Commun.}, vol.~24, no.~2, pp. 98--105, 2016.

\bibitem{thilina2013machine}
K.~M. Thilina, K.~W. Choi, N.~Saquib, and E.~Hossain, ``Machine learning
  techniques for cooperative spectrum sensing in cognitive radio networks,''
  \emph{IEEE J. Sel. Areas Commun.}, vol.~31, no.~11, pp. 2209--2221, 2013.

\bibitem{zhuang2020machine}
Z.~Zhuang, L.~Xu, J.~Li, J.~Hu, L.~Sun, F.~Shu, and J.~Wang,
  ``Machine-learning-based high-resolution doa measurement and robust
  directional modulation for hybrid analog-digital massive mimo transceiver,''
  \emph{Science China Information Sciences}, vol.~63, no.~8, pp. 1--18, 2020.

\bibitem{shu2021spatial}
F.~Shu, L.~Liu, L.~Yang, X.~Jiang, G.~Xia, Y.~Wu, X.~Wang, S.~Jin, J.~Wang, and
  X.~You, ``Spatial modulation: an attractive secure solution to future
  wireless network,'' \emph{arXiv preprint arXiv:2103.04051}, 2021.

\bibitem{jie2021high}
Q.~Jie, X.~Zhan, F.~Shu, Y.~Ding, B.~Shi, Y.~Li, and J.~Wang,
  ``High-performance passive eigen-model-based detectors of single emitter
  using massive mimo receivers,'' \emph{arXiv preprint arXiv:2108.02011}, 2021.

\bibitem{zhang2021direction}
R.~Zhang, B.~Shim, and W.~Wu, ``Direction-of-arrival estimation for large
  antenna arrays with hybrid analog and digital architectures,'' \emph{IEEE
  Trans. Signal Process.}, vol.~70, pp. 72--88, 2021.

\bibitem{chen2008stochastic}
C.~E. Chen, F.~Lorenzelli, R.~E. Hudson, and K.~Yao, ``Stochastic
  maximum-likelihood doa estimation in the presence of unknown nonuniform
  noise,'' \emph{IEEE Trans. Signal Process.}, vol.~56, no.~7, pp. 3038--3044,
  2008.

\bibitem{chiani2014distribution}
M.~Chiani, ``Distribution of the largest eigenvalue for real wishart and
  gaussian random matrices and a simple approximation for the tracy--widom
  distribution,'' \emph{Journal of Multivariate Analysis}, vol. 129, pp.
  69--81, 2014.

\bibitem{hagan1997neural}
M.~T. Hagan, H.~B. Demuth, and M.~Beale, \emph{Neural network design}.\hskip
  1em plus 0.5em minus 0.4em\relax PWS Publishing Co., 1997.

\bibitem{prahofer2004exact}
M.~Pr{\"a}hofer and H.~Spohn, ``Exact scaling functions for one-dimensional
  stationary kpz growth,'' \emph{Journal of statistical physics}, vol. 115,
  no.~1, pp. 255--279, 2004.

\bibitem{akaike1998information}
H.~Akaike, ``Information theory and an extension of the maximum likelihood
  principle,'' in \emph{Selected papers of hirotugu akaike}.\hskip 1em plus
  0.5em minus 0.4em\relax Springer, 1998, pp. 199--213.

\end{thebibliography}
\end{document}